\documentclass[sigconf,nonacm,natbib=false]{acmart}
\makeatletter
\global\@ACM@balancefalse
\makeatother
\usepackage{balance}

\setcopyright{none}
\renewcommand\footnotetextcopyrightpermission[1]{}
\usepackage[
  backend=biber,
  style=numeric,
  sorting=none,
  giveninits=true,
  maxbibnames=3,
  minbibnames=3,
  doi=false,
  isbn=false,
  url=false,
  eprint=false
]{biblatex}
\DeclareNameAlias{author}{family-given}
\DeclareNameAlias{editor}{family-given}

\renewrobustcmd*{\bibinitperiod}{}
\renewrobustcmd*{\bibinitdelim}{}
\DeclareDelimFormat{multinamedelim}{\addcomma\space}
\DeclareDelimFormat{finalnamedelim}{\addcomma\space}
\DefineBibliographyStrings{english}{andothers={et\addabbrvspace al\adddot}}
\DeclareFieldFormat[article,inproceedings,incollection,thesis,misc]{title}{#1}
\DeclareFieldFormat{pages}{#1}

\AtBeginBibliography{\sloppy}

\newbibmacro*{compact:author-title}{%
  \printnames{author}%
  \setunit{\addperiod\space}%
  \printfield{title}}

\DeclareBibliographyDriver{article}{%
  \usebibmacro{bibindex}%
  \usebibmacro{begentry}%
  \usebibmacro{compact:author-title}%
  \newunit\newblock
  \printfield{journaltitle}%
  \setunit{\addcomma\space}%
  \printfield{year}%
  \iffieldundef{volume}
    {}
    {\setunit{\addcomma\space}%
     \printfield{volume}%
     \iffieldundef{number}{}{\printtext{\mkbibparens{\printfield{number}}}}}%
  \iffieldundef{pages}{}{\setunit{\addcolon\space}\printfield{pages}}%
  \usebibmacro{finentry}}

\DeclareBibliographyDriver{inproceedings}{%
  \usebibmacro{bibindex}%
  \usebibmacro{begentry}%
  \usebibmacro{compact:author-title}%
  \newunit\newblock
  \printfield{booktitle}%
  \setunit{\addcomma\space}%
  \printfield{year}%
  \iffieldundef{pages}{}{\setunit{\addcolon\space}\printfield{pages}}%
  \usebibmacro{finentry}}

\DeclareBibliographyDriver{thesis}{%
  \usebibmacro{bibindex}%
  \usebibmacro{begentry}%
  \usebibmacro{compact:author-title}%
  \newunit\newblock
  \printtext{Master's thesis}%
  \iflistundef{institution}{}{\setunit{\addcomma\space}\printlist{institution}}%
  \setunit{\addcomma\space}%
  \printfield{year}%
  \usebibmacro{finentry}}

\DeclareBibliographyDriver{misc}{%
  \usebibmacro{bibindex}%
  \usebibmacro{begentry}%
  \usebibmacro{compact:author-title}%
  \iffieldundef{howpublished}{}{\setunit{\addperiod\space}\printfield{howpublished}}%
  \iffieldundef{year}{}{\setunit{\addcomma\space}\printfield{year}}%
  \usebibmacro{finentry}}

\title{Cross-Media Scientific Research Achievements Query Based on Ranking Learning}

\author{Benzhi Wang}
\affiliation{%
  \institution{School of Computer Science (National Pilot School of Software Engineering), Beijing University of Posts and Telecommunications; Beijing Key Laboratory of Intelligent Telecommunication Software and Multimedia}
  \city{Beijing}
  \country{China}}

\author{Meiyu Liang}
\authornote{Corresponding author.}
\affiliation{%
  \institution{School of Computer Science (National Pilot School of Software Engineering), Beijing University of Posts and Telecommunications; Beijing Key Laboratory of Intelligent Telecommunication Software and Multimedia}
  \city{Beijing}
  \country{China}}

\author{Ang Li}
\affiliation{%
  \institution{School of Computer Science (National Pilot School of Software Engineering), Beijing University of Posts and Telecommunications; Beijing Key Laboratory of Intelligent Telecommunication Software and Multimedia}
  \city{Beijing}
  \country{China}}

\begin{abstract}
With the advent of the information age, the scale of data on the Internet is getting larger and larger, and it is full of text, images, videos, and other information. Different from social media data and news data, scientific research achievement information has the characteristics of many proper nouns and strong ambiguity. The traditional single-mode query method based on keywords can no longer meet the needs of scientific researchers and managers of the Ministry of Science and Technology. Scientific research project information and scientific research scholar information contain a large amount of valuable scientific research achievement information. Evaluating the output capability of scientific research projects and scientific research teams can effectively assist managers in decision-making. In view of the above background, this paper expounds on the research status from four aspects: characteristic learning of scientific research results, cross-media research results query, ranking learning of scientific research results, and cross-media scientific research achievement query systems.
\end{abstract}

\keywords{science and technology big data, cross-media retrieval, cross-media semantic association learning, deep language model, semantic similarity}

\begin{document}
\maketitle

The scale of scientific research results has grown rapidly with the progress of the times and is now enormous. Major universities and research institutions produce scientific research results continuously. These results may come from a professor, a student, or a team, and may be published individually or as part of a scientific research project. Scientific research results include a variety of scientific and technological resource information in different media, including images and text. Efficiently collecting, processing, and storing such multi-source and heterogeneous cross-media scientific research data is an important issue~\cite{shi2019deepCollaborative}. Deep modularity-based community detection can help identify cohesive collaboration groups in such research networks~\cite{yang2016modularityCommunity}.

With the advent of the information age, traditional query systems that retrieve scientific research results only through keywords have gradually lagged behind current needs. For researchers, text-result query services such as CNKI are relatively complete, but matching models based only on keywords can no longer satisfy daily retrieval needs. Synonymy and polysemy cannot be ignored in text retrieval, and simple keyword matching cannot solve these problems. Deep language models such as BERT provide a foundation for addressing polysemy. Adjacent work on multi-view clustering and uncertainty-aware network estimation further illustrates the challenges of combining heterogeneous representations~\cite{xue2019deepLowRank,li2017varianceConstrained}. Using deep language models to retrain scientific research results for query needs therefore has profound practical and research significance. At the same time, single-modal retrieval will be gradually replaced by cross-modal retrieval. Researchers may want to find relevant papers and patents through a circuit diagram or a neural-network model diagram.

For project managers at universities, funding committees, the Ministry of Science and Technology, and the National Natural Science Foundation of China, existing systems mainly support keyword-based exact queries of scientific research projects. Information must be queried and counted individually according to a person's name, which is inconvenient. Query technology that integrates scientific research results, scientific research teams, scientific research projects, and related information is intended to meet these application scenarios. Supervisors urgently need technical tools to obtain valuable information from large collections of scientific research scholars and teams. Interpretable machine-learning models can make the resulting intelligent decisions more transparent to managers~\cite{li2019interpretableDecision}.

\section{Characteristic Learning of Scientific Research Results}

Cross-media scientific research results contain scientific and technological resources from different fields and modalities. Existing studies propose different methods for converting cross-media data into unified features~\cite{gong2019semiSupervisedFeatureMapping,chen2016onlineCommodityCrossMedia,wu2020neuralImageRetrieval,qi2019semiSupervisedGraphFeature,sun2020compressedCNNRetrieval}. Teacher-student distillation for graphs with incomplete features and structure provides a related mechanism for recovering information before representation learning~\cite{huo2023t2gnn}. The main approach processes data from each modality according to its characteristics and uses cross-media cooperative learning to map the data into a unified feature subspace. The cross-media data considered here mainly include text and images. Establishing a unified feature subspace requires converting the text and images in cross-media scientific research results into effective, unified feature vectors.

Computers are not good at directly processing sets of textual symbols. Converting text into feature vectors is therefore indispensable in natural-language-processing tasks. The simplest methods are one-hot encoding and TF-IDF. One-hot encoding~\cite{rodriguez2018beyondOneHot} uses an $N$-dimensional vector to indicate whether a term is one of $N$ words, with exactly one bit set to 1. TF-IDF~\cite{kim2019multiCoTraining} is a statistical method for estimating the importance of a word in a document or file. A word's importance is proportional to its frequency in an individual document and decreases as its overall corpus frequency increases. Although these two methods have novel applications~\cite{xu2020oneHotPatent}, their shortcomings are evident. They do not consider symbol order or positional information and depend heavily on the corpus. Heterogeneous graph attention provides an alternative for semi-supervised classification when scientific resource descriptions are short and sparsely labeled~\cite{hu2019heterogeneousGAT}. Directly applying one-hot and TF-IDF encoding to scientific research results is insufficient because the number of scientific entities and proper nouns is enormous, the resulting vectors become excessively long, and neither method handles synonyms well.

In 2013, Google open-sourced the Word2vec word-vector calculation tool. Word2vec has been used for topic modeling, recommendation, and document embedding~\cite{park2020ldaWord2vecTravel,caselles2018word2vecRecommendation,wu2018wordMoversEmbedding}. It can be trained on large datasets and million-word dictionaries, and distances between word embeddings can measure similarity. Word2vec has two basic implementations: CBOW and Skip-gram. CBOW predicts the current word from its context~\cite{liu2020cbowSentiment,novak2020cbowTag}, whereas directional Skip-gram predicts context from the current word~\cite{song2018directionalSkipGram,li2017distributedKalman}. These approaches encode contextual and positional information, but Word2vec assumes that word semantics are determined by frequent context. Scientific and technological big data contain polysemous words such as ``nuclear'' and ``apple.'' ``Nuclear'' may refer to a nucleus in physics, a processor core in computing, or a kernel function in artificial intelligence. Context-based regression and local-density estimation have likewise been used to identify anomalous representations~\cite{hu2020anomalyDetection}, illustrating the importance of modeling context rather than isolated terms.

In 2018, Google announced BERT, which performed well on several natural-language-processing tasks. Transformer self-attention provides the architectural basis~\cite{vaswani2017attention}, while self-supervised and semi-supervised learning can learn representations from data without complete human annotation~\cite{zhai2019s4l}. BERT is trained on a massive corpus and can provide transferable feature representations~\cite{devlin2018bert,chao2019bertDst}. Retrieval-oriented masked-autoencoder pretraining further adapts language representations to search tasks~\cite{xiao2022retroMAE}. BERT output can be used as word embeddings for downstream tasks and either fine-tuned or fixed as a feature extractor. The source code and models were released publicly~\cite{google2018bertRepo}. Although such models can be applied directly to scientific and technological big data, they do not adapt perfectly to accurate scientific-research-result queries. General-domain training provides limited recognition of specialized entities and proper nouns, so further domain training is required.

An image is represented in a computer as a sequence of pixels and can be more difficult to interpret than text. Traditional feature-extraction methods include the Scale-Invariant Feature Transform (SIFT)~\cite{xiao2022siftUav} and Histograms of Oriented Gradients (HOG)~\cite{wang2021hogLicensePlate}. These methods express image features through prior knowledge and are interpretable, but they are designed for specific tasks. SIFT is suitable for image matching and three-dimensional modeling, HOG for pedestrian or object detection, and local binary patterns for face recognition and image classification. They cannot directly form mappings to text in other modalities.

Most current image-recognition tasks use machine learning~\cite{zheng2019deepImageReview}, and convolutional neural networks are a popular direction~\cite{tang2018conditionalGANRecognition,li2017recursiveStateA}. Zhang et al.~\cite{zhang2022handwrittenTransformer} proposed a handwritten English text-recognition method based on a convolutional neural network and Transformer. Transformer replaces recurrent LSTM processing with global self-attention, and its combination with CNNs supports segmentation-free recognition. VGGNet repeatedly stacks convolution and pooling layers to extract deep image features. A VGGNet-based method has been used to classify and recognize the main organs of tomato plants with data augmentation~\cite{zhou2017tomatoVGG}. Omnidirectional image super-resolution with bi-projection fusion further illustrates how task-specific geometric reconstruction can enhance image representations~\cite{wang2024omnidirectionalSuperResolution}. Like text feature extractors, image feature extractors must be optimized for specific tasks. General models such as BERT and VGGNet perform well on ordinary text and images but require further optimization for accurate scientific research achievement queries.

\section{Cross-Media Research Results Query}

The key issue in cross-media scientific research result queries is extracting same-dimensional information from different modalities~\cite{li2022crossMediaRetrieval}. For example, the text term ``polymer'' is related to a picture of epoxy resin because epoxy resin is a kind of polymer. Cross-media query methods map resources of different modalities into a common feature subspace and calculate similarities within that subspace~\cite{liu2019coupledDictionary,qi2019dictionaryCrossMedia,duan2019cnnCrossMedia,zong2019l2pCrossMedia}. When source data cannot be centralized, federated supervised cross-modal retrieval can learn aligned representations while limiting direct data exchange~\cite{li2024federatedCrossModal}. Canonical correlation analysis is a traditional feature-mapping algorithm that learns a linear relationship between two sets of variables~\cite{zhuang2012sparseCCA}.

Deep cross-modal retrieval can use correspondence autoencoders to model common correlations between modalities and to optimize a joint objective~\cite{feng2015deepCrossModal}. Other work has matched Chinese poetry and folk music through emotional characteristics, using emotional similarity to connect text and music~\cite{xing2020crossMediaSynesthesia}. Resource-oriented library cross-media knowledge services place construction and management of cross-media knowledge graphs at their core and use cross-media discovery to support innovation~\cite{liu2020libraryKnowledgeService}. Cross-media social-network security topic mining has explored topics from the rich media of Sina Weibo with deep learning~\cite{zhou2020socialSecurityTopic}, while reinforcement learning, adversarial learning, and semantic analysis support cross-media social search~\cite{kou2016socialSearch}. Federated graph neural networks extend decentralized representation learning to cross-graph node classification~\cite{guan2021federatedGNN}. Reinforcement-based active client selection can further improve participation decisions under heterogeneous graph distributions~\cite{wang2025reinforcementClientSelection}.

Existing cross-media query systems do not fully integrate cross-media information with deep semantics. They often rely on keywords, clustering~\cite{sun2009knn}, or topic mining based on cross-media similarity. Research on cross-media scientific achievement queries is therefore not yet mature, and effectively learning semantic information across modalities remains an urgent problem.

\section{Ranking Learning of Scientific Research Results}

With the growth of big data, manual sorting and scoring are no longer suitable after information retrieval. Modern web-page ranking considers many factors, making manual scoring impractical. Applying machine learning to ranking has led to learning-to-rank methods~\cite{wang2018positionBias,ai2019groupwiseScoring,hu2019unbiasedLambdaMART}. PageRank and HITS are classical ranking algorithms~\cite{yu2020pagerankCrawler}, but they do not directly incorporate user behavior. Learning to rank can automatically optimize a ranking model from feedback and support personalized ranking~\cite{pereira2020rankviz,zhu2020pointwiseLTR}. Generative recommendation models can also unify retrieval and ranking within a single generation process~\cite{zhang2025unifiedGenerativeRecommendation}.

Learning-to-rank methods mainly take three forms: single-document PointWise methods, document-pair PairWise methods~\cite{xiong2017pairwiseReview,yu2020enhancedFM}, and document-list ListWise methods~\cite{gong2018listwiseLoss}. PointWise methods transform ranking into classification or regression. PairWise methods process query-document pairs and order documents by relative correlation. ListWise methods learn a scoring function over an entire list. Topic-similarity-weighted voting can further improve ListWise ranking accuracy~\cite{liu2016topicSimilarityRank}.

Another ranking framework combines matrix decomposition, clustering, and deep neural networks~\cite{yang2018matrixFactorizationRank}. Network-state modeling illustrates how complex dependencies can be represented~\cite{li2017recursiveStateB}, while image-fusion research demonstrates feature integration across structured signals~\cite{xu2013imageFusion}. Graph-based modeling can also improve abstractive multi-document summarization~\cite{li2020graphSummarization}, showing how document relationships can refine representations. Semantic-similarity attention combined with hypergraph convolution captures higher-order relations among scientific publications~\cite{li2026hypergraphPublication}. Deep interest networks mine historical user behavior, weight it through attention, and support personalized retrieval and click-through-rate prediction~\cite{zhou2018deepInterestNetwork}. Filter-enhanced MLP models provide an efficient alternative for encoding ordered interaction signals in sequential recommendation~\cite{zhou2022filterMLP}.

Convolutional models can further process feature vectors for ranking. A multi-channel convolutional document-list model performs list-level reranking~\cite{cao2020listCNN}. Related representation-learning and consensus models illustrate the broader use of deep transformations and coordinated optimization~\cite{fang2020cycleGAN,lin2009averageConsensus}. Online learning-to-rank methods balance speed and quality by learning from user interactions~\cite{oosterhuis2017onlineLTR,zhao2017slidingConsensus}. Self-supervised graph co-training for session-based recommendation offers another way to stabilize sequential representations through complementary graph views~\cite{xia2021graphCoTraining}. Complex models are more expressive but require more interactions and computation, whereas simple models train quickly but may converge to suboptimal solutions. Cascaded designs seek a balance between rapid learning and high-quality convergence.

\section{Cross-Media Scientific Research Achievement Query System}

With the continuing growth of the Internet and scientific innovation, the number of scientific research achievements increases each year. Traditional databases can no longer provide sufficiently efficient retrieval for massive result collections~\cite{yang2015ontologyRetrieval}. Lucene is a full-text search engine based on an inverted index~\cite{sha2019lucene,sun2021searchableEncryption}. It includes indexing, searching, and management modules, analyzes large document collections, divides them into terms, and constructs inverted index tables~\cite{ding2015luceneSorting,du2019ciphertextRetrieval}. Lucene ranks documents using score functions and index structures~\cite{xing2018secureIndex,quan2020searchEngine}.

Multi-source index configuration supports large document-search systems~\cite{liao2020documentSearch}, and Boolean query methods can be implemented over multi-source Lucene indexes~\cite{qiu2018multiSourceLucene}. When standalone Lucene is insufficient, distributed search engines such as Solr~\cite{li2021solr} and Elasticsearch~\cite{liu2021elasticSearch} provide distributed indexing, failover, and load balancing. For decentralized information networks, FedSIN uses federated self-adaptive learning to obtain representations without centralizing all data~\cite{li2026fedSIN}. Communication-efficient reinforcement federated learning can further reduce coordination costs through dynamic client selection and adaptive gradient compression~\cite{pan2025rfcsc}. Search-index signals can also be combined for prediction and recommendation~\cite{yuan2020webSearchIndex}.

Existing scientific research achievement query systems, including the National Science and Technology Achievement Information Service System~\cite{nstasSystem}, support keyword queries and filtering by application industry, source, and completion time. Research on decoupling under velocity-varying conditions provides a general systems perspective on separating interacting factors, although it is not itself an information-retrieval method~\cite{li2014lpvControl}. Multi-view scholar clustering with dynamic interest tracking can represent both multiple research perspectives and changes in a scholar's interests over time~\cite{li2023scholarClustering}. AMiner is a large-scale scientific and technological information-mining platform that analyzes researchers, literature, and academic activities~\cite{aminerPlatform}. AMiner includes large collections of researchers, knowledge concepts, papers, and citation relationships and makes data available to researchers. However, its functions are not specifically targeted at integrating a scholar's research projects with longitudinal statistical summaries of that scholar's work.

\section{Conclusion}

Existing cross-media query systems do not sufficiently integrate cross-media information and deep semantics. They mainly search by keywords or perform clustering and topic mining based on cross-media similarity. Research on querying cross-media scientific research achievements is still immature. Effectively learning semantic information across modalities and solving ranking problems after retrieval therefore remain urgent tasks.

\section{Acknowledgements}

This work was supported by the National Key R\&D Program of China (2018YFB1402600) and the National Natural Science Foundation of China (61772083, 61877006, 61802028, and 62002027).

\balance
\printbibliography[title={References}]

\end{document}